\documentclass[aps,prl,twocolumn,superscriptaddress]{revtex4}

\usepackage{graphicx}
\usepackage{graphicx}
\usepackage{here}
\begin{document}
\title{{\it In-situ} photoemission study of Pr$_{1-x}$Ca$_x$MnO$_3$ 
epitaxial thin films with suppressed charge fluctuations}

\author{H. Wadati}
\email{wadati@phas.ubc.ca}
\homepage{http://www.geocities.jp/qxbqd097/index2.htm}
\altaffiliation[Present address: ]
{Department of Physics and Astronomy, 
University of British Columbia, Vancouver, 
British Columbia V6T-1Z1, Canada}
\affiliation{Department of Physics, University of Tokyo,
Bunkyo-ku, Tokyo 113-0033, Japan}

\author{A. Maniwa}
\affiliation{Department of Applied Chemistry, University of Tokyo, 
Bunkyo-ku, Tokyo 113-8656, Japan}

\author{A. Chikamatsu}
\affiliation{Department of Applied Chemistry, University of Tokyo, 
Bunkyo-ku, Tokyo 113-8656, Japan}

\author{I. Ohkubo}
\affiliation{Department of Applied Chemistry, University of Tokyo, 
Bunkyo-ku, Tokyo 113-8656, Japan}

\author{H. Kumigashira}
\affiliation{Department of Applied Chemistry, University of Tokyo, 
Bunkyo-ku, Tokyo 113-8656, Japan}

\author{M. Oshima}
\affiliation{Department of Applied Chemistry, University of Tokyo, 
Bunkyo-ku, Tokyo 113-8656, Japan}

\author{A. Fujimori}
\affiliation{Department of Physics, University of Tokyo,
Bunkyo-ku, Tokyo 113-0033, Japan}

\author{M. Lippmaa}
\affiliation{Institute for Solid State Physics, University of Tokyo, 
Kashiwa, Chiba 277-8581, Japan}

\author{M. Kawasaki}
\affiliation{Institute for Materials Research, Tohoku University, 
2-1-1 Katahira, Aoba, Sendai 980-8577, Japan}

\author{H. Koinuma}
\affiliation{National Institute for Materials Science, 1-2-1 Sengen, Tsukuba, Ibaraki 305-0047}

\date{\today}
\begin{abstract}
We have 
performed an {\it in-situ} photoemission 
study of Pr$_{1-x}$Ca$_x$MnO$_3$ (PCMO) thin films 
grown on LaAlO$_3$ (001) substrates and observed the 
effect of epitaxial strain 
on the electronic structure. 
We found that the 
chemical potential shifted monotonically 
with doping, unlike bulk PCMO, 
implying the disappearance of 
incommensurate charge fluctuations of bulk PCMO. 
In the valence-band spectra, we found 
a doping-induced energy shift toward 
the Fermi level ($E_F$) but there was no 
spectral weight transfer, which was observed in bulk PCMO. 
The gap at $E_F$ was clearly seen in the experimental 
band dispersions determined by angle-resolved 
photoemission spectroscopy and 
could not be explained 
by the metallic band structure of the 
C-type antiferromagnetic state, probably due to 
localization of electrons along the ferromagnetic 
chain direction or due to another type of spin-orbital 
ordering.  
\end{abstract}
\maketitle
Strongly correlated systems have 
attracted great interest because of 
their various 
interesting physical properties, 
such as metal-insulator 
transition, colossal magnetoresistance (CMR), 
and the ordering 
of spin, charge, and orbitals \cite{rev}. 
One of the peculiar features of these systems 
is their high sensitivity to external stimuli. 
Pressure as well as carrier concentration 
are among the most important parameters. The effects of carrier 
doping and chemical pressure on their electronic structures 
have been extensively studied by 
transport measurements and various spectroscopic methods, 
including photoemission spectroscopy, although 
it has been impossible to 
directly observe the electronic structures under 
physical hydrostatic 
pressure by photoemission spectroscopy 
because of the fundamental limitation of 
the technique. 
However, 
if one grows thin films 
epitaxially on single crystalline 
substrates, 
one can effectively perform photoemission measurements 
under (anisotropic) high pressure. 
As for the high-$T_c$ cuprates, 
Abrecht {\it et al.} \cite{LSCOfilm,LSCOfilm2} 
performed an {\it in-situ} 
angle-resolved photoemission spectroscopy (ARPES) 
study of La$_{2-x}$Sr$_x$CuO$_4$ (LSCO) thin films 
grown on SrLaAlO$_4$ (001) substrates 
(under in-plane compressive strain), and 
found that the topology of the 
Fermi surface changed from 
that of unstrained bulk LSCO 
concomitantly with an increase of $T_c$ 
from 38 K to 40 K. 

Effects of pressure should be particularly 
striking for charge-orbital-related phenomena 
since the charge and orbital degrees of freedom 
are strongly coupled to lattice distortion. 
In this sense, direct observation of pressure 
effects on the electronic structure of hole-doped 
perovskite manganites is highly attractive. 
The hole-doped perovskite manganites 
$R_{1-x}A_x$MnO$_3$, where 
$R$ is a rare-earth ($R=$ La, Nd, Pr) and 
$A$ is an alkaline-earth atom ($A=$ Sr, Ba, Ca), 
exhibit CMR and spin, charge, and orbital
ordering \cite{RamirezMn,
RaoMn,PrellierMn,Hungry,TokuraMn}. 
Most of the half-doped manganites ($x\simeq 0.5$) 
with a small bandwidth $W$ exhibit the so-called ``CE-type'' 
antiferromagnetic (AF) charge ordering (CO) 
with alternating 
Mn$^{3+}$ and Mn$^{4+}$ states within 
the (001) plane in the form of stripes \cite{Jirak}. 
The inter-stripe 
distance increases with further hole doping 
and such charge modulations persist at high 
temperatures as fluctuations well above the 
CO temperature $T_{CO}$ \cite{little,Shimomura}. 
Pr$_{1-x}$Ca$_x$MnO$_3$ (PCMO), 
where $W$ is the smallest, 
has a particularly stable CO state 
at low temperatures in a wide 
hole concentration range $0.3 \le x \le 0.75$ 
\cite{tomiokaco}. 
It is known that the magnetic and electronic phases 
of Mn oxides can be controlled in thin films 
grown on substrates with various lattice parameters. 
For example, La$_{0.5}$Sr$_{0.5}$MnO$_3$ thin films remain 
ferromagnetic (FM) 
on (LaAlO$_3$)$_{0.3}$-(SrAl$_{0.5}$Ta$_{0.5}$O$_3$)$_{0.7}$ 
(LSAT) substrates, but becomes A-type AF 
on SrTiO$_3$ (STO) substrates, and 
C-type AF on LAO 
substrates \cite{Konishi}. 
The CO states of (Nd$_{1-x}$Pr$_x$)$_{0.5}$Sr$_{0.5}$MnO$_3$ 
thin films were also found to be controlled by the strain 
effects from the substrates \cite{NPSMO}. 
Recently, it was reported that 
PCMO thin films grown on STO $(001)$ substrates 
have higher CO temperatures than bulk samples, indicating 
that in-plane tensile strain leads to the stabilization 
of the CO state \cite{PCMOSTO}. 

In this work, we have studied the electronic structure of 
PCMO thin films grown 
on LaAlO$_3$ (LAO) substrates 
by photoemission spectroscopy. 
This work is the first experimental observation of 
the electronic structure of the strained 
manganites. 
The fabricated PCMO thin films were under 
compressive strain from the LAO substrates, 
which is considered to suppress CO or charge 
modulation \cite{PCMOSTO}. 
From core-level 
photoemission studies, we found that the chemical 
potential shift as a function of hole doping 
was not suppressed in the doping region 
where incommensurate charge fluctuations are 
observed in the bulk samples. 
From the valence-band spectra, we found 
that no new states appeared near 
the Fermi level ($E_F$) with hole 
doping. These results are in striking contrast to 
the recent results of PCMO bulk samples reported 
by Ebata {\it et al.} \cite{EbataPCMO}, 
who concluded 
that the chemical potential pinning and the 
spectral weight transfer do occur, 
and are considered to be 
spectroscopic evidence for the suppression of 
incommensurate charge modulation in PCMO thin films grown on LAO. 

\begin{figure}
\begin{center}
\includegraphics[width=9cm]{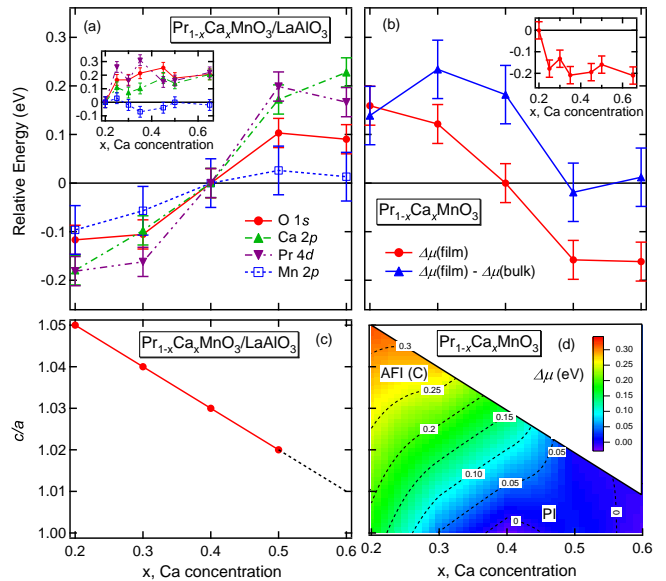}
\caption{(Color online) 
Core-level binding-energy and chemical-potential 
shifts in PCMO. 
(a) Shift of each core level of PCMO thin films grown 
on LAO substrates. 
(b) Chemical potential shift ($\Delta \mu$(film)) 
deduced from the core-level shifts 
and the difference of the chemical potential 
($\Delta\mu$(film) $-$ $\Delta\mu$(bulk)) 
between the film and the bulk. 
The insets to (a) and (b) 
show the results of bulk samples taken 
from Ref.~\cite{EbataPCMO}. 
(c) $c/a$ of PCMO thin films as a function of $x$ taken 
from Ref.~\cite{ManiwaPCMO}. 
(d) $\Delta\mu$ as a function of $c/a$ and $x$. PI and AFI (C) 
denote the paramagnetic insulating and C-type 
antiferromagnetic insulating states, respectively.}
\label{shift1pc}
\end{center}
\end{figure}

The experiments were performed at beamlines 
1C and 2C of the Photon Factory, 
High Energy Accelerators Research 
Organization (KEK), using a combined laser 
MBE photoemission spectrometer 
system \cite{Horiba}. 
Epitaxial thin films 
of PCMO with a thickness of about 400 $\mbox{\AA}$ 
were fabricated by the pulsed 
laser deposition method 
from ceramic targets of 
desired chemical compositions. 
Single crystals of LAO (001) 
were used as the substrates. 
Atomically flat 
step-and-terrace structures were observed by 
atomic force microscopy. 
The crystal structure was characterized 
by four circle x-ray diffraction measurements. 
The in-plane lattice constants of 
the PCMO thin films were the same as that of 
LAO ($a=3.792$ $\mbox{\AA}$), 
confirming the epitaxial and coherent growth 
of the thin films on the substrates. 
For all the compositions, 
the out-of-plane lattice constants 
were longer than the 
in-plane lattice constants, indicating that 
the present PCMO thin films were under 
compressive strain [see Fig.~\ref{shift1pc} (c)]. 
In the 
LEED patterns, sharp $1\times 1$ spots were observed without 
surface-reconstruction-derived spots. 
The electrical resistivities were high and 
showed no jump as a function of temperature, indicative of a 
suppression of CO and the associated lattice 
distortion due to the compressive strain 
imposed by the LAO substrates. 
Details of 
the sample growth and characterization can 
be found in Ref.~\cite{ManiwaPCMO}. 
The photoemission spectra were taken using
a Gammadata Scienta SES-100 spectrometer. 
All the spectra were measured at room temperature, 
except for ARPES measurements (20 K). 
The total energy resolution was about 
$150-400$ meV depending on photon energies. 
The Fermi level ($E_F$) position was 
determined by measuring gold spectra. 

From the core-level photoemission 
spectra, 
we found that all the spectra 
were shifted toward lower binding 
energies with $x$, as plotted in Fig.~\ref{shift1pc} (a). 
Here, the ``relative energies'' are referenced to 
the core levels of the $x=0.4$ sample. 
The chemical potential shift $\Delta\mu$ 
can be obtained from 
the average of the energy shifts of 
the O $1s$, Ca $2p$, and Pr $4d$ 
core levels as in the case of bulk PCMO \cite{EbataPCMO}. 
Figure \ref{shift1pc} (b) shows $\Delta\mu$ thus determined 
plotted as a function of $x$. The shift of the chemical potential 
is monotonic without any sign of suppression at least up to $x=0.5$, 
which is similar to bulk and thin film LSMO 
\cite{Matsuno2,horibaLSMO} but 
is quite different from bulk PCMO \cite{EbataPCMO}. 
The suppression of the chemical potential shift has 
been observed in the region of incommensurate charge 
fluctuations \cite{little,Shimomura} in 
bulk PCMO ($x>0.3$) at room temperature 
\cite{EbataPCMO}, 
La$_{2-x}$Sr$_x$NiO$_4$ \cite{SatakeLSNO}, 
and underdoped La$_{2-x}$Sr$_x$CuO$_4$ \cite{Ino}. 
Their origin has been attributed to 
dynamical stripe-type charge fluctuations, 
a kind of ``microscopic 
phase separation'' 
between hole-rich and hole-poor regions. 
That is, the suppression of the chemical potential 
shift occurs when the distance between 
stripes changes as a function of 
hole concentration. 
Such a ``microscopic phase separation'' may be 
absent in the PCMO thin films 
considering the suppression of charge modulation 
caused by the compressive 
strain effects from the LAO substrates. 
Figure \ref{shift1pc} (b) shows 
$\Delta\mu$(film) $-$ $\Delta\mu$(bulk), plotted 
as a function of $x$. Note 
that $c/a$ extrapolates to 1 at 
$x\sim 0.7$ in PCMO thin films 
on LAO substrates \cite{ManiwaPCMO} 
[see Fig.~\ref{shift1pc} (c)]. Chemical 
potential shifts to higher energy by applying compressive 
strain. 
Figure \ref{shift1pc} (d) shows interpolated 
$\Delta\mu$ as a function of $c/a$ and 
$x$. 
This panel clearly shows the pinning of $\Delta\mu$ in the 
paramagnetic insulating region 
near $c/a\sim 1$ for $x\ge 0.3$, 
whereas the pinning disappears for 
smaller $x$ and larger $c/a$. 
 
\begin{figure}
\begin{center}
\includegraphics[width=8.5cm]{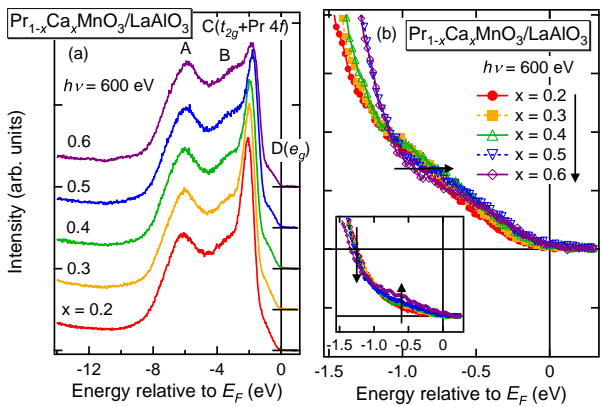}
\caption{(Color online) Doping dependence of the 
valence-band photoemission spectra 
of PCMO thin films. 
(a) Valence-band photoemission spectra 
in a wide energy range. 
(b) Valence-band spectra near $E_F$. 
The inset shows the result of bulk samples taken 
from Ref.~\cite{EbataPCMO}. 
(c) Valence-band spectra near $E_F$ with energy positions 
shifted considering the chemical potential shift. 
Arrows indicate systematic spectral feature shifts 
with increasing doping.}
\label{val1pc}
\end{center}
\end{figure}

Figure \ref{val1pc} (a) shows the doping 
dependence of the valence-band 
photoemission spectra. 
Following Ref.~\cite{EbataPCMO}, 
structures A, B, C, and D 
are assigned to 
Mn $3d$ - O $2p$ bonding, 
non-bonding O $2p$, 
Mn $3d$ $t_{2g}$ plus Pr $4f$, and 
Mn $3d$ $e_g$ states, respectively. 
A gap (absence of finite density of states 
at $E_F$) was seen for all values of $x$, 
which is considered to 
be a natural consequence of the insulating 
nature of the PCMO 
thin films. Structures A-D moved toward $E_F$ 
upon hole doping. This is contrasted with 
the results of bulk PCMO, where 
spectral weight transfer occurs 
near $E_F$ from high 
to low energies with hole doping 
as shown in the inset to 
Fig.~\ref{val1pc} (b) \cite{EbataPCMO}. 
Sekiyama 
{\it et al.} \cite{SekiyamaNSMO2} 
also reported finite intensity at $E_F$ 
in the insulating state of Nd$_{1-x}$Sr$_x$MnO$_3$. 
The main panel of Fig.~\ref{val1pc} (b) shows the 
valence-band spectra of the PCMO thin films 
near $E_F$. 
The overall shift of the spectra is 
attributed to the 
effect of the chemical potential shift, and 
no new states appeared near $E_F$ with hole doping. 
From the present spectra near $E_F$, 
we conclude that our PCMO thin films were 
good insulators without any 
dynamical ``phase separation''. 

The absence of 
chemical potential ``pinning'' and 
spectral weight transfer toward $E_F$ 
in the thin films, 
which were observed in bulk PCMO \cite{EbataPCMO}, 
are attributed to the suppression of incommensurate 
charge fluctuations 
due to the compressive strain effects 
from the LAO substrates, and are considered as 
the spectroscopic evidence for the change of 
the electronic structures due to the epitaxial 
strain effects from the substrates. 

\begin{figure}
\begin{center}
\includegraphics[width=8.5cm]{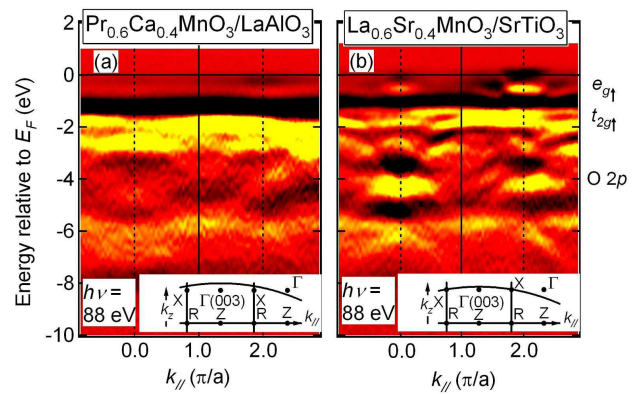}
\caption{(Color online) ARPES spectra of PCMO (a) and 
LSMO (taken from Ref.~\cite{Chika}) (b) 
taken at 88 eV. Bright parts 
correspond to energy bands. 
The insets show the traces in $k$ space.}
\label{ARPES1}
\end{center}
\end{figure}

Figure \ref{ARPES1} (a) shows the 
ARPES spectra of a PCMO ($x=0.4$) 
thin film taken with a photon energy of 88 eV. 
Here, the second derivatives 
of the energy distribution curves 
are plotted as a false-color image, where bright 
parts correspond to peaks or shoulders in 
energy distribution curves. 
Hereafter, $k_{\parallel}$ 
denotes the in-plane momentum 
expressed in units of $\pi/a$. 
The structures at $-(1.5-3)$ eV show 
weak dispersions and are assigned to the Mn $3d$ 
$e_{g\uparrow}$ and $t_{2g\uparrow}$ bands. 
The structures at $-(3-7)$ eV show strong dispersions 
and are assigned to the O $2p$ bands. 
There is 
no intensity at $E_F$, consistent with the 
insulating behavior of this film. 
The same plot for an LSMO 
thin film ($x=0.4$) grown on an STO substrate 
is shown for comparison 
in Fig.~\ref{ARPES1} (b) \cite{Chika}. 
The dispersions of the O $2p$ bands were 
similar between these two spectra, while 
those of the Mn $3d$ bands were very different. 
The Mn $3d$ $e_{g\uparrow}$ bands 
show a clear dispersion and cross $E_F$ 
in LSMO, but are weak and 
show only very weak dispersions in PCMO.  
 
\begin{figure}
\begin{center}
\includegraphics[width=8.5cm]{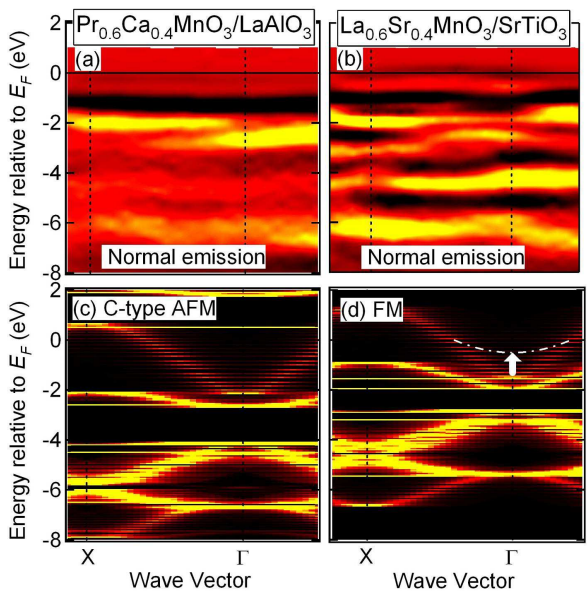}
\caption{(Color online) 
Comparison of the ARPES spectra 
measured in normal emission geometry with 
tight-binding band-structure 
calculation (obtained with 
the parameter set of 
$\epsilon_d-\epsilon_p=2.0$ eV, 
$(pd\sigma)=-1.9$ eV, and 
the exchange splitting $\Delta E = 4.8$ eV.) 
Bright parts correspond to energy bands. 
(a) Experimental band structure of PCMO ($x=0.4$). 
(b) Experimental band structure of LSMO ($x=0.4$) 
taken from Ref.~\cite{ChikamatsuJES}. 
(c) Tight-binding calculation of the C-type AF state. 
(d) Tight-binding calculation of the FM state. The arrow 
indicates the effect of mass renormalization. 
Both simulations have taken into account the finite 
photoelectron mean-free path \cite{Wadatiphase}. 
In (a) and (c) we plotted the photoemission spectral 
weight calculated by projecting 
the obtained C-type AF bands 
to their original PM bands \cite{fold}.}
\label{ARPES2}
\end{center}
\end{figure}

The magnetic ground state of the PCMO ($x=0.4$) 
thin film may 
be inferred from the lattice constants and the 
phase diagram of LSMO proposed by Konishi {\it et al.} 
\cite{Konishi}. 
The $c/a$ of the PCMO ($x=0.4$) films 
was 1.03 as shown in Fig.~\ref{shift1pc} (c). 
From the phase diagram 
in Ref.~\cite{Konishi}, 
$c/a=1.03$ and $x=0.4$ is just at the boundary of 
the FM and C-type AF states. 
We confirmed that this film was not FM by 
magnetization measurements, and considered 
it to be in the C-type AF state. 
In the C-type AF state, 
FM metallic chains are formed along 
the c-axis. Therefore, we have performed 
normal emission ARPES measurements to 
study the out-of-plane band dispersions 
as shown in Fig.~\ref{ARPES2} (a). 
For comparison, Fig.~\ref{ARPES2} (b) 
shows the 
normal-emission ARPES spectra of 
an LSMO ($x=0.4$) thin film taken from 
Ref.~\cite{ChikamatsuJES}. Here again, 
the dispersions of the O $2p$ bands are 
similar between these two spectra. 
On the other hand, 
the dispersions of the Mn $3d$ $e_{g\uparrow}$ bands 
are clear in LSMO (b), 
but very weak in PCMO (a). 
In order to interpret the 
experimental band dispersions, we have performed a 
tight-binding (TB) band-structure calculation 
with empirical parameters. 
Here, we have performed the calculation by 
assuming 
the C-type AF state for PCMO and 
the FM state for LSMO as shown in 
Figs.~\ref{ARPES2} (c) and (d). 
In the case of LSMO, 
agreement between 
experiment and calculation is good 
except for the 
narrowing of the conduction band 
due to strong electron correlation. 
In the case of PCMO, 
the strong dispersion 
of the Mn $3d$ $e_{g\uparrow}$ bands 
predicted by 
the calculation of Fig.~\ref{ARPES2} (c) 
was not observed in the experiment of 
Fig.~\ref{ARPES2} (a). 
The $e_g$ electrons 
appear to be 
localized along the ferromagnetic chain direction, 
probably due to disorder and/or electron correlation. 
This result is also consistent with the fact that 
in C-type AF materials the expected one-dimensional 
metalicity has not been observed so far. 

In summary, we have performed 
an {\it in-situ} photoemission study of 
Pr$_{1-x}$Ca$_x$MnO$_3$ (PCMO) thin films grown 
on LaAlO$_3$ (LAO) substrates. 
From the core-level 
photoemission study, we found that unlike bulk PCMO, in strained films 
the chemical potential shifted monotonically with doping, 
implying the disappearance of incommensurate 
charge fluctuations. 
In the valence-band spectra, we found 
no spectral weight at the Fermi level ($E_F$) nor 
doping-induced spectral weight transfer toward 
$E_F$, also unlike bulk PCMO. 
The gap at $E_F$ was clearly seen in the experimental 
band dispersions determined by angle-resolved 
photoemission spectroscopy (ARPES) and 
could not be explained 
by the metallic band structure of the 
C-type AF state 
as previously proposed, probably due to 
the localization of electrons along the ferro-chain 
direction caused by disorder and electron 
correlation. 

Informative discussion with K.~Ebata and 
experimental support by M.~Takizawa are gratefully 
acknowledged. 
This work was supported by a Grant-in-Aid 
for Scientific Research (A16204024) from 
the Japan Society for the Promotion of 
Science (JSPS) and a Grant-in-Aid 
for Scientific Research in Priority Areas 
``Invention of Anomalous Quantum Materials'' 
from the Ministry of Education, Culture, 
Sports, Science and Technology. 
H. W. acknowledges financial support from JSPS. 
The work 
was done under the approval of the Photon Factory 
Program Advisory Committee (Proposal 
Nos.~2005G101 and 2005S2-002) 
at the Institute of Material Structure Science, KEK. 

\bibliography{LVO1tex}
\end{document}